\documentstyle[twoside,fleqn,miami]{article}

\newcommand{\AmS}{{\protect\the\textfont2
  A\kern-.1667em\lower.5ex\hbox{M}\kern-.125emS}}

\title{\hspace{3.5mm}{\bf Superconductivity and Stoichiometry in the}\\
\hspace{3.5mm}{\bf BSCCO-family Materials}}

\author{\hspace{3.5mm}{\bf Marshall Onellion}\thanks{Department of Physics,
University of Wisconsin, Madison, WI 53706.}}

\begin{document}

\begin{abstract}
\hspace{48mm}{\it Received 31 January 1995}

\hspace{48mm}{\bf \large ----------------------------------------%
--------------------------------}

\hspace{48mm}We report on magnetization, c-axis and ab-plane resistivity,
critical current, elec-

\hspace{48mm}troni band structure and superconducting gap properties.
Bulk measurements and

\hspace{48mm}photoemission data were taken on similar samples.

\hspace{48mm}{\bf \large ----------------------------------------%
--------------------------------}

\hspace{48mm}{\bf KEY WORDS:} Resistivity; band structure; superconducting
gap; symmetry.
\end{abstract}

\maketitle

\vspace{5mm}
\hspace{8mm}I have two purposes in writing this report. One is to provide
experimental data that has been reproduced in different laboratories, and
so can be viewed as reliable, to serve as the basis for theoretical models.
The other is to argue that there is a close connection between changes
observed in the superconducting properties (gap size and temperature
dependence, critical current, magnetic field dependence) and normal state
properties (symmetry of electronic states, topology of Fermi surface, c-axis
resistivity).

\vspace{5mm}
\hspace{8mm}In our studies, we have found c-axis resistivity data versus oxygen
content that are consistent with and extend earlier reports.[1-4] Note that
the c-axis resistivity is reduced by a factor of as much as $\times 250$ for
overdoped compared to underdoped samples. However, neither we nor our
colleagues [1-4] have yet incorporated sufficient oxygen to observe fully
metallic $(dR/dT > 0)$ behavior for all temperatures down to the
superconducting transition temperature ($T_c$), as has been observed for
YBCO-123.[5]  The significant point to these measurements is that there is a
marked change in the c-axis resistivity, and hence interlayer coupling,
with oxygen stoichiometry.

\vspace{5mm}
\hspace{8mm} For the same samples for which we measured c-axis resistivity, we
also performed angle-resolved photoemission measurements of the normal
state.[4] The issue was to determine whether samples that exhibited the
change in c-axis resistivity also exhibited any differences in their Fermi
surfaces. We have reported elsewhere [6] that there is a finite
interlayer coupling between the adjacent $CuO_2$ planes of the bilayer.
We chose Pb-doped BSCCO-2212 samples. By performing careful TEM
measurements,[4]
we determined that the overdoped and underdoped samples were isostructural in
the $CuO_2$ (ab-)planes. The only structural difference is a change in the
periodicity of the superlattice modulation, with the overdoped samples
exhibiting a larger period. We found that there is a change of symmetry in
the normal state quasiparticle band states as the amount of oxygen is
changed. Specifically, the electronic states that comprise the Fermi surface
in the $k_x = \pm k_y$ directions exhibited a change in
symmetry with oxygen stoichiometry, while the symmetry of the states in the
$k_x$ and $k_y$ directions was unaffected by the oxygen stoichiometry.
Our results indicate one of two possibilities. One is that the c-axis
(interlayer coupling) affects the symmetry of the electronic states. The
other possibility is that the electronic states arise due to many-body
effects.[7] If the electronic states are interpreted as many-body states,[7]
preliminary calculations indicate that interactions beyond the
nearest-neighbor must be included to explain the data.[7]

\vspace{5mm}
\hspace{8mm}In addition to the symmetry of the normal state electronic states,
our data indicate two other important points about the \  Fermi \ surface.
We \  find

\newpage
\pagestyle{myheadings}
\markboth{{\bf \small Onellion}}{{\bf \small Superconductivity and
Stoichiometry in BSCCO-family Materials.}}
$\!\!\!\!\!\!$that there is an \  extended \  van Hove \  singularity,[8]
consistent with other reports.[9-11] The extent of the singularity is
reduced for overdoped samples.[4] A pocket, absent for underdoped samples,
develops around the $(\pi,0)$ point for overdoped samples,[4,8] consistent
with earlier work by C. Olson and colleagues.[12,13] Further, it is noteworthy
that P. Aebi and colleagues have used our samples to study the presence and
strength of the shadow bands they reported earlier.[14] In their previous
work,[14] they reported observing shadow bands only on samples that were
particularly flat and well-ordered (to reduce scattering that averages over
the Brillouin zone). Using our samples, they find such shadow bands, and find
that the shadow bands are weaker, but present, for the overdoped samples
($T_c$ = 75 K) compared to the underdoped samples ($T_c$ = 80 K).[15] These
data indicate that both the symmetry of the Fermi surface electronic states,
and the topology of the Fermi surface, change with oxygen doping.

\vspace{5mm}
\hspace{8mm} Note that we have found results consistent with other
investigators
for those measurements where consistency is expected. For that reason, our
results as to the symmetry and topology of the Fermi surface electronic states
appear representative of samples in different laboratories. As an additional
check of our samples, we have studied the current versus voltage measurements
for supercurrent along the c-axis and applied magnetic field in the
ab-plane.[16] The issue is whether the change in c-axis resistivity, and
Fermi surface electronic states, both normal state properties, are related to
superconducting state properties.

\vspace{5mm}
\hspace{8mm}We found that for our samples, the $CuO_2$ planes are stacked very
close to parallel, with an ab-plane misalignment below 0.02 degrees. This
allowed us to orient the applied magnetic field parallel to the ab-plane with
high accuracy. The results at lower magnetic fields (below 1 Tesla) reproduce
earlier work by Kleiner et.al.[17,18] For underdoped samples, the data
indicate a S-I-S Josephson junction stacking of planes along the c-axis. For
overdoped samples, the critical current density increases between
$(\times 100 -\times 1000)$ and the stacking is S-N(S')-S. Our results
indicate a marked increase in critical current density for overdoped samples,
the same samples that exhibit the changes in normal state properties.

\vspace{5mm}
\hspace{8mm}However, neither our results [16] nor those of our colleagues,
[17,18] conclusively establish whether the c-axis coupling in overdoped
samples is Josephson-junction or three dimensional. One new, and pertinent,
result is that overdoped samples obey the Kim-Stevens relation,[19] as do
three-
dimensional superconductors. We found $H_{c_2}$ is already 14 Tesla only
4 K below $T_c$.
[16]

\vspace{5mm}
\hspace{8mm}In magnetization studies of cuprate superconductors, much has
been made of the ``fishtail'' behavior,[20-22] a region of applied magnetic
field for which the magnetization increases with increasing applied field.
Using samples similar to those used in photoemission studies,
X.Y. Cai {\it et al.}[22] reported that the fishtail exists for both
underdoped and overdoped samples, although such samples exhibit very
different quantitative magnetization response. The data were interpreted as
indicating that there are always stronger and weaker superconducting regions.
These are the same sample types described above.
Consequently, the fact that all samples exhibit the fishtail behavior indicates
that all samples exhibit both stronger and weaker superconducting regions.

\vspace{95mm}

{\small
$\!\!\!\!\!${\bf Fig. 1.} Photoemission spectra in the normal state along the
$k_ x = k_y$ direction for He-annealed, Pb-doped sample (gap = 0-2 meV) and
oxygen-annealed, overdoped sample (gap = 10-12 meV).[27]
}

\vspace{5mm}
\hspace{8mm}For the same sample types, we also conducted angle-resolved
photoemission measurements of the size of the superconducting gap
for two high-symmetry directions, $k_x$ (Cu-O-Cu bond axis in real space) and
$k_x = k_y$ (Bi-O-Bi bond axis perpendicular to superlattice
modulation).[23,24] Figure One illustrates the results. We
found that for samples with less oxygen, particularly if underdoped, the
gap along $k_x = k_y$ is indistinguishable from zero ($2 \pm 2$ meV),
consistent with earlier reports by B.O. Wells and Z.-X. Shen.[25,26]
However, for overdoped samples, the data indicate unambiguously that the
superconducting gap in the $k_x = k_y$ direction is non-zero.[27] Observing a
non-zero gap in the $k_x = k_y$ direction has also been reported earlier by
several research groups.[28-31] One significant point is that this study is
the first to directly connect the size of the superconducting gap and the
bulk critical current and magnetization properties of a cuprate
superconductor system.

\vspace{95mm}

{\small
$\!\!\!\!\!${\bf Fig. 2.} Superconducting gap (in meV) versus temperature
for $k_x$ direction (diamonds) and $k_x = k_y$ direction (squares).[32]
}

\vspace{5mm}
\hspace{8mm}These results can be regarded as solid, because they either
reproduce earlier work or have been independently confirmed by other
investigators, including other investigators using our samples.[15] The last
experimental result that we wish to report is the temperature and
momentum-resolved study of the superconducting gap for oxygen overdoped
samples.[32] We studied the variation of the superconducting gap with
temperature for two Brillouin zone directions, $k_x$ and $k_x = k_y$.
Figure Two illustrates the results. We found that the temperature
dependence of the superconducting gap along the two symmetry directions is
qualitatively different. In particular, the gap along the $k_x = k_y$
direction becomes indistinguishable from zero at $0.82T_c$, while along
$k_x$ the gap remains at 90-100\% of its value at $0.35T_c$. Some parts of
the data in Fig. 2 have been independently confirmed. The rapid, non-BCS,
increase of the gap with decreasing temperature in the $k_x$ direction has
been confirmed by J.C. Campuzano and colleagues.[33] The non-zero gap in the
$k_x = k_y$ direction at lower temperatures has been reported earlier.[28-31]

\vspace{5mm}
\hspace{8mm}What inferences can be drawn that are based strictly on data
reproduced in different laboratories? I am confident of the following:
\begin{itemize}
\item Saying that the symmetry of the order parameter (gap) remains identical
for all stoichiometry has been ruled out;[27-29,31-34]

\item There is a marked increase in interlayer coupling for overdoped samples,
as
reflected in bulk measurements [1-4,16-18] and the symmetry of
the normal state electronic states;[6]

\item The shadow bands reported earlier by P. Aebi et.al. [14] are observed for
samples from different laboratories, and exist well into the overdoped
regime;[15]

\item The gap increases more rapidly for temperatures below $T_c$ than a
would a BCS superconductor.[32,33]
\end{itemize}

\vspace{5mm}
\hspace{8mm}In addition, if all the results of of Ref. 32 are confirmed by
independent work, these results indicate that, for overdoped samples, just
below $T_c$ there may be only d-wave pairing, while at lower temperatures a
more complicated pairing interaction exists. However, for underdoped samples,
the small value of the gap along the $k_x = k_y$ direction (at $0.35T_c$)
indicates that a predominant d-wave pairing extends to lower temperature.

\newpage
{\bf ACKNOWLEDGEMENTS}
\vspace{7mm}

\hspace{8mm}Various parts of the work reported were performed in collaboration
with, above all, my students and postdocs, Jian Ma, Ronald Kelley and Christoph
Quitmann, as well as with other colleagues, including Cai Xueyu, Yi
Feng, David Larbalestier, Philippe Almeras, Helmut Berger, and Giorgio
Margaritondo. Financial support was provided by the U.S. NSF, Ecole
Polytechnique F\'{e}d\'{e}rale, Fonds National Suisse de la Recherche
Scientifique, and the Deutsche Forschungsgemeinschaft.

\section*{References} 
{\small

\begin{enumerate}
\item C. Kendziora {\it et al., Phys. Rev. B} {\bf 48}, 3531 (1993).

\item D. Mandrus {\it et al., Phys. Rev. B} {\bf 44}, 2418 (1991).

\item F.X. Reiz {\it et al., IEEE Trans. Appl. Supercon.} {\bf 3}, 1190 (1993).

\item Jian Ma {\it et al., Phys. Rev. B}, in press (1995).

\item L. Forro {\it et al.}, Phys. Rev. B {\bf 46}, 6626 (1992).

\item C. Quitmann {\it et al.}, submitted to {\it Phys. Rev. B}.

\item E. Dagotto {\it et al., Phys. Rev. Lett.} {\bf 73}, 728 (1994).

\item Jian Ma {\it et al., Phys. Rev. B}, in press (Feb. 1, 1995).

\item A. Abrikosov {\it et al., Physica C} {\bf 214}, 73 (1993).

\item K. Gofron {\it et al., Phys. Rev. Lett.} {\bf 73}, 3302 (1994).

\item D. Dessau {\it et al., Phys. Rev. Lett.} {\bf 71}, 2781 (1993).

\item C. Olson {\it et al., Phys. Rev. B} {\bf 42}, 381 (1990).

\item B.O. Wells {\it et al., Phys. Rev. Lett.} {\bf 65}, 3056 (1990).

\item P. Aebi {\it et al., Phys. Rev. Lett.} {\bf 72}, 2757 (1994).

\item P. Aebi {\it et al}, unpublished.

\item X.Y. Cai {\it et al}, unpublished.

\item R. Kleiner and P. Mueller, {\it Phys. Rev. B} {\bf 49}, 1327 (1994).

\item R. Kleiner {\it et al., Phys. Rev. B} {\bf 50}, 3942 (1994);
R. Kleiner, op. cit., 6919 (1994).

\item R.D. Parks, Ed. {\it ``Superconductivity''}, , Marcel Dekker, NY,
1969, Ch. 19.

\item M. Daeumling {\it et al., Nature} {\bf 346}, 332 (1990).

\item V.N. Kopylov {\it et al, Physica C} {\bf 170}, 291 (1990).

\item X.Y. Cai {\it et al., Phys. Rev. B} {\bf 50}, 16774 (1994).

\item S. Sunshine {\it et al., Phys. Rev. B} {\bf 38}, 893 (1988).

\item M.D. Kirk {\it et al., Science} {\bf 242}, 1673 (1988).

\item B.O. Wells {\it et al., Phys. Rev. B} {\bf 46}, 11830 (1992).

\item Z.-X. Shen {\it et al., Phys. Rev. Lett.} {\bf 70}, 1553 (1993).

\item R.J. Kelley {\it et al., Phys. Rev. B} {\bf 50}, 590 (1994) and
unpublished.

\item C. Olson {\it et al., Science} {\bf 245}, 731 (1989);
{\it Solid State Commun.} {\bf 76}, 411 (1990).

\item  R.J. Kelley {\it et al., Phys. Rev. Lett.} {\bf71}, 4051 (1993).

\item H. Ding {\it et al., Phys. Rev. B} {\bf 50}, 1333 (1994).

\item H. Ding {\it et al.}, preprint and private communication.

\item Jian Ma {\it et al., Science}, in press (1995).

\item J.C. Campuzano, private communication and oral presentation at the
1994 Wisconsin Synchrotron Radiation Center Users Meeting.

\item A. Abrikosov, private communication.
\end{enumerate}
}

\end{document}